\documentclass{jfm}

\usepackage{graphicx}
\usepackage{natbib}
\usepackage{upmath}

\ifCUPmtlplainloaded \else
  \checkfont{eurm10}
  \iffontfound
    \IfFileExists{upmath.sty}
      {\typeout{^^JFound AMS Euler Roman fonts on the system,
                   using the 'upmath' package.^^J}%
       \usepackage{upmath}}
      {\typeout{^^JFound AMS Euler Roman fonts on the system, but you
                   dont seem to have the}%
       \typeout{'upmath' package installed. JFM.cls can take advantage
                 of these fonts,^^Jif you use 'upmath' package.^^J}%
      }
  \else
  \fi
\fi

\ifCUPmtlplainloaded \else
  \checkfont{msam10}
  \iffontfound
    \IfFileExists{amssymb.sty}
      {\typeout{^^JFound AMS Symbol fonts on the system, using the
                'amssymb' package.^^J}%
       \usepackage{amssymb}%
         \let\leq=\leqslant
         
      }{}
  \fi
\fi

\ifCUPmtlplainloaded \else
  \IfFileExists{amsbsy.sty}
    {\typeout{^^JFound the 'amsbsy' package on the system, using it.^^J}%
     \usepackage{amsbsy}}
    {\providecommand\boldsymbol[1]{\mbox{\boldmath $##1$}}}
\fi

\newsavebox{\astrutbox}
\sbox{\astrutbox}{\rule[-5pt]{0pt}{20pt}}

\newcommand\etal{\mbox{\textit{et al.}}}

\newcommand\eg{e.g.\ }

\title[Helical structure of longitudinal vortices in turbulent wall-bounded flow]{Helical structure of longitudinal vortices embedded in turbulent wall-bounded flow}

\author[C. M. Velte, M. O. L. Hansen and V. L. Okulov]%
{C\ls L\ls A\ls R\ls A\ns M.\ns V\ls E\ls L\ls T\ls E,\ns M\ls A\ls
R\ls T\ls I\ls N\ns O.\ns L.\ns H\ls A\ls N\ls S\ls E\ls N\ls\break
\and V\ls A\ls L\ls E\ls R\ls Y\ns L.\ns O\ls K\ls U\ls L\ls O\ls V}

\affiliation{Department of Mechanical Engineering, Technical
University of Denmark, DK-2800 Kgs. Lyngby, Denmark}

\pubyear{2008} \volume{...} \pagerange{...--...}
\date{...}

\begin{document}

\maketitle

\begin{abstract}
Embedded vortices in turbulent wall-bounded flow over a flat plate,
generated by a passive rectangular vane-type vortex generator with
variable angle $\beta$ to the incoming flow in a low-Reynolds number
flow ($Re=2600$ based on the inlet grid mesh size $L=0.039\;$m and
free stream velocity $U_{\infty} = 1.0\;$m\,s$^{-1}$) have been
studied with respect to helical symmetry. The studies were carried
out in a low-speed closed-circuit wind tunnel utilizing Stereoscopic
Particle Image Velocimetry (SPIV). The vortices have been shown to
possess helical symmetry, allowing the flow to be described in a
simple fashion. Iso-contour maps of axial vorticity revealed a
dominant primary vortex and a weaker secondary one for $20^{\circ}
\leq \beta \leq 40^{\circ}$. For angles outside of this range, the
helical symmetry was impaired due to the emergence of additional
flow effects. A model describing the flow has been utilized, showing
strong concurrence with the measurements, even though the model is
decoupled from external flow processes that could perturb the
helical symmetry. The pitch, vortex core size, circulation and the
advection velocity of the vortex all vary linearly with the device
angle $\beta$. This is important for flow control, since one thereby
can determine the axial velocity induced by the helical vortex as
well as the swirl redistributing the axial velocity component for a
given device angle $\beta$. This also simplifies theoretical
studies, \eg to understand and predict the stability of the vortex
and to model the flow numerically.
\end{abstract}


\section{Introduction}
Streamwise vortices embedded in turbulent boundary layers is a
common phenomenon and is seen \eg in the treatment of free organized
structures \cite[see e.g.][and references therein]{Adrian2007},
G\"{o}rtler vortices in boundary layers over walls of streamwise
concave curvature \cite[see][]{Gortler1955}, corner vortices with an
axial velocity component, vortex rings near walls and as horseshoe
vortices folding around objects attached to a wall
\cite[][]{Adrian2007}. Often longitudinal vortices are generated
with passive devices called vortex generators. A vortex generator is
similar to a wing with a small aspect ratio mounted normally to a
surface with an angle of incidence to the oncoming flow. It is
designed to overturn the boundary layer flow via large scale
motions, thereby redistributing the streamwise momentum in the
boundary layer which aids in preventing separation. Vortex
generators were formally introduced by H. D. Taylor
\cite[see][]{Taylor1947} as an aid in suppressing separation in
diffusers. Many studies have presented (nominal) guidelines for
optimizing the effect of forced mixing for these passive devices for
varying geometries and flow conditions, \cite[see
\eg][]{Schubauer1960,Pearcey,Godard2006}. Further, a review on
low-profile vortex generators was written by \cite{Lin2002}. The
applicability of controlled near-wall vortices in engineering is
vast, since vortices can transport both heat and momentum, aiding in
cooling or re-energizing the lowest part of the boundary layer.
Being able to control/optimize parameters such as the strength and
size of the longitudinal vortices to the existing flow setting is
highly desired and it is therefore of interest to develop theories
and models which can predict and describe these. Some models have
been proposed in order to describe the flow, both theoretically
\cite[see \eg][]{Smith1994} as well as computationally \cite[see
\eg][]{Liu1996,You2006}. The model of \cite{Smith1994} predicts the
flow field induced by low-profile triangular vanes (extending
approximately to the logarithmic region of the boundary layer) in a
zero pressure gradient boundary layer. The method modifies the
governing equations based on the scales of the geometry and the
oncoming flow. Good agreement is found with experiments, however,
this model only treats low-profile devices extending to a fraction
of the boundary layer height. Having a similar geometric
configuration, Liu \etal\, (1996) introduced vortices numerically
using body forces and utilized the fact that the azimuthal velocity
distribution of the device-induced vortices is similar to that of
Lamb-Oseen vortices. The non-uniform axial component was obtained by
introducing a Gaussian distributed streamwise force component.
However, this was merely introduced and never motivated more than on
a purely empirical basis to compensate for the momentum deficit in
the wake of the device.

The main objective of this work is the experimental investigation of
device-generated vortices to define helical vortex structures in
wall-bounded flow and to create a new model which more correctly can
describe the vortex flow. Previously, a lot of experimental work was
done describing embedded vortices in boundary layer flows using
single point measurement techniques \cite[see \eg Schubauer \etal\,
1960;][]{Shabaka1985}. However, the development of Stereoscopic
Particle Image Velocimetry (SPIV) allows non-intrusive instantaneous
measurement realizations of the flow in a plane and is the
predominating measurement technique for these investigations today
\cite[see \eg Godard \etal\, 2006;][]{Velte2008}. SPIV measurements
in spanwise planes downstream of a single rectangular vortex
generator on a flat plate have been conducted and investigated. This
configuration is subject to a parametric study, investigating the
effect on the helical vortex when varying the angle of the actuating
device to the incoming flow. A turbulent boundary layer profile was
considered suitable due to a fuller velocity profile. This also
makes the results applicable to flows at more realistic Reynolds
numbers. The turbulence level was generated using an inlet grid to
yield a high enough turbulence intensity to obtain a turbulent
boundary layer profile. Results show that the vortex generator gives
rise to longitudinal vortices that possess helical symmetry. A
simple theoretical flow model is put forward based on the hypothesis
of helical symmetry of the generated vortices and the Gaussian
distribution of the vorticity field. The axial and azimuthal
vorticity components are coupled according to the definition for
helical symmetry of vorticity fields; $\omega_r = 0$ and
$\omega_{\theta}/\omega_{z}=r/l$, where $l$ represents the helical
pitch, see figure \ref{fig:helix}($b$). Even though the vortex
generators operated in a turbulent boundary layer, yielding
relatively large perturbations, the vortex was observed to be stable
in the experiments. None of the previous work has dealt with the
helical symmetry of embedded longitudinal vortices and specifically,
the longitudinal vortices generated by vortex generators have not
previously been known to possess helical symmetry.


\section{Experimental method}

\begin{figure}
\begin{minipage}[t]{0.50\linewidth}
\includegraphics[width=7.0cm]{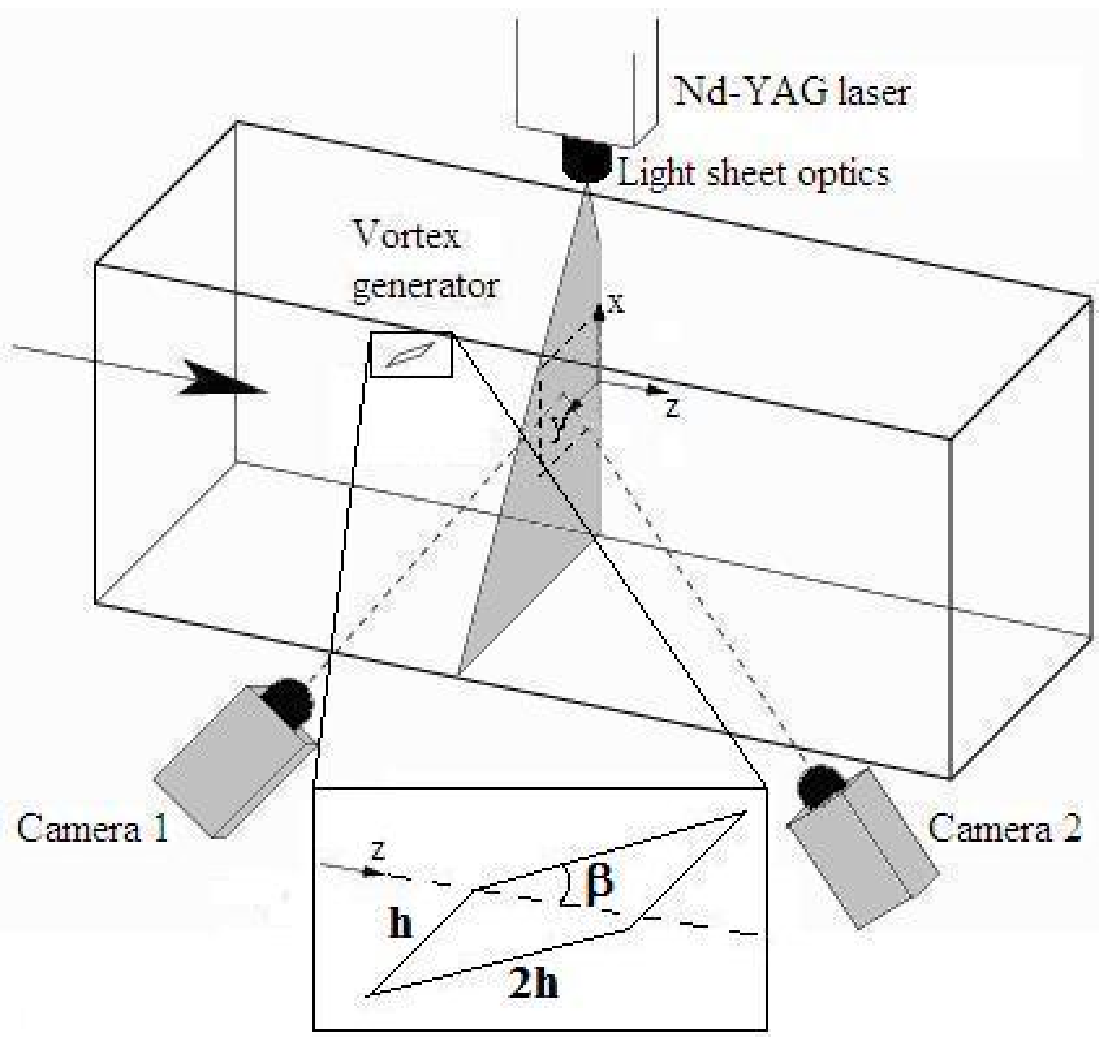}
\caption{Schematic of the experimental set-up and device geometry.
The large arrow to the left indicates the main flow direction and
$\beta$ the device angle. The measurement plane in the laser sheet
has been indicated by dashed lines.} \label{fig:setup}
\end{minipage}\hspace{0.6cm}
\begin{minipage}[t]{0.45\linewidth}
\includegraphics[width=6.0cm]{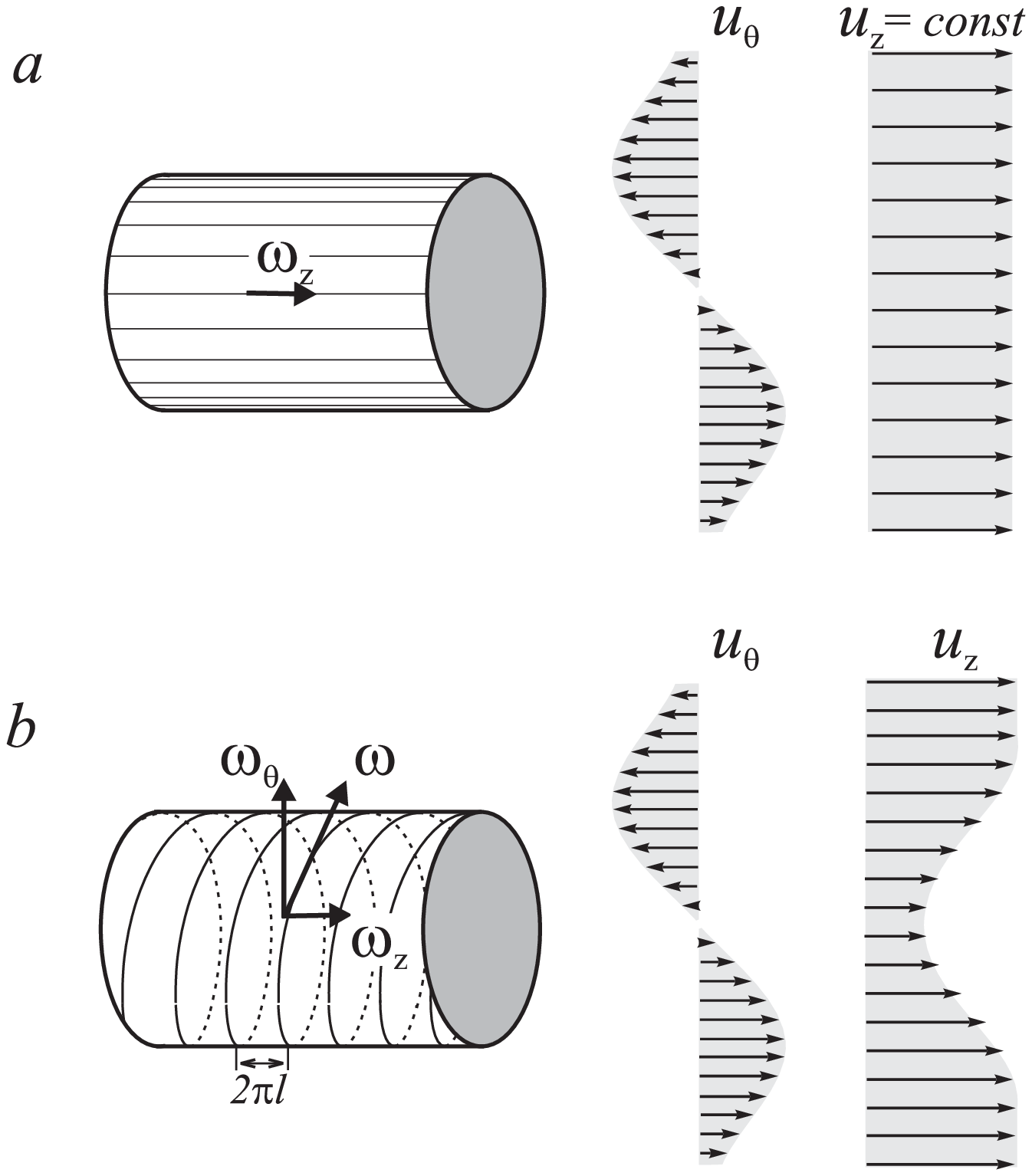}
\caption{Sketch of vorticity field and induced velocity profile by
Lamb-Oseen vortex with rectilinear vortex lines (a) and Batchelor
vortex with helical structure of vortex lines (b).}
\label{fig:helix}
\end{minipage}
\end{figure}

Consider the test section setup in figure \ref{fig:setup}. The
measurements were carried out in a closed-circuit wind tunnel with
an 8:1 contraction ratio and a test section of cross-sectional area
300\,$\times$\,600$\;$mm with length 2$\;$m. At the inlet of the
test section, a turbulence-generating grid with mesh length 39$\;$mm
was situated. The test section had optical access through the top
and bottom walls as well as through the sidewall opposite to the
wall with the attached vortex generator. The coordinate system is
defined in figure \ref{fig:setup}. $z$ is the axial flow direction,
$y$ is the wall-normal direction and $x$ is the spanwise direction.

The experiments were conducted at a free stream velocity of
$U_{\infty}=1.0\;$m\,s$^{-1}$. The wind tunnel speed was obtained by
measuring the pressure drop across an orifice plate. The turbulence
intensity at the inlet has from LDA measurements been found to be
13\%. The boundary layer thickness at the position of the vortex
generator has been estimated from LDA measurements to be
approximately $\delta_{VG}=25\;$mm. The actuator, as seen in figure
\ref{fig:setup}, is a rectangular vane of the same height as the
local boundary layer thickness, $h=\delta_{VG}$, with a length of
$2h$. The vortex generator was positioned on a vertical wall in the
centre of the test section with its trailing edge 750$\;$mm
downstream of the inlet grid when at zero angle to the mean flow. In
order to easily and accurately alter the device angle, the vortex
generator was attached to a pin which could be accessed from outside
of the test section through a hole in the test section wall. This
pin was in turn attached to a pointer arm placed over a protractor
indicating the relative angle of the actuator to the mean flow
direction. The protractor had a radius of 200$\;$mm and grading for
integer values of each degree. The device angle of incidence $\beta$
could therefore be determined with a relatively high accuracy. The
measurements were conducted in a spanwise plane, with plane normal
parallel to the test section walls, positioned five device heights
downstream of the vortex generator. The measurement plane has been
indicated by a dashed line in figure \ref{fig:setup}. Measurements
were conducted for 5$^{\circ} \leq \beta \leq$ 85$^{\circ}$ with
5$^{\circ}$ angle spacings.

The stereoscopic PIV equipment was mounted on a rigid stand and
included a double cavity NewWave Solo 120XT Nd-YAG laser (wavelength
532$\;$nm), capable of delivering light pulses of 120$\;$mJ. The
pulse width, i.e. the duration of each illumination pulse, was
10$\;$ns. The light sheet thickness at the measurement position was
2$\;$mm and was created using a combination of a spherical convex
and a cylindrical concave lens. The equipment also included two
Dantec Dynamics HiSense MkII cameras (1344\,$\times$\,1024 pixels)
equipped with 60$\;$mm lenses and filters designed to only pass
light with wavelengths close to that of the laser light. Both
cameras were mounted on Scheimpflug angle adjustable mountings. The
seeding, consisting of DEHS (Di-Ethyl-Hexyl-Sebacin-Esther) droplets
with a diameter of 2--3$\;\umu$m, was added to the flow downstream
of the test section in the closed-circuit wind tunnel in order to
facilitate a homogeneous distribution of the particles before they
enter the test section. The laser was placed above the test section,
illuminating a plane normal to the test section walls, see figure
\ref{fig:setup}. The two cameras were placed in the forward
scattering direction. The angle of each respective camera to the
laser sheet was 45$^{\circ}$. The f-numbers of the cameras were set
to 2.8, yielding a depth of field which is small but sufficient to
cover the thickness of the laser sheet and keeping all illuminated
particles in focus while still attaining sufficient scattered light
from the tracer particles. In order to avoid reflections from the
wall and the vortex generator within the wavelength band of the
camera filters, these areas were treated with a fluorescent dye,
Rhodamine 6G, mixed with matt varnish to obtain a smooth surface and
to ensure that the dye stayed attached. A calibration target was
aligned with the laser sheet. This target had a well defined
pattern, which could be registered by the two cameras to obtain the
geometrical information required for reconstructing the velocity
vectors received from each camera to obtain a full description of
all three velocity components in the plane. Calibration images were
recorded with both cameras at five well defined streamwise positions
throughout the depth of the laser sheet in order to capture the
out-of-plane component in the reconstructed coordinate system of the
measurement plane under consideration. A linear transform was
applied to these images for each camera respectively to perform the
reconstruction. This procedure was executed both previous to and
after the conduction of the measurements to ensure that no drift had
occurred. The images were processed using Dantec Dynamicstudio
software version 2.0. Adaptive correlation was applied using
refinement with an interrogation area size of 32\,$\times$\,32
pixels. Local median validation was used in the immediate vicinity
of each interrogation area to remove spurious vectors between each
refinement step. The overlap between interrogation areas was 50\%.
For each measurement position, 500 realizations were acquired. The
recording of image maps was done with an acquisition rate of
1.0$\;$Hz, ensuring statistically independent realizations based on
the convection velocity $U_{\infty}=1.0\;$m\,s$^{-1}$ and the mesh
size $d=0.039\;$m, yielding a time scale of
$t=d/U_{\infty}=0.039\;$s. The velocity vector maps contain 73 by 61
vectors. The linear dimensions of the interrogation areas
(x,y)$=$(1.55,1.04) mm can be compared to the Taylor microscale and
the Kolmogorov length scale estimated to $\lambda_f \approx$ 9$\;$mm
and $\eta \approx\;$0.5 mm from Laser Doppler Anemometry (LDA)
measurements \cite[]{Schmidt1997}.


\section{Modelling of the longitudinal vortex}\label{sec:theory}

\noindent The existence of Lamb-Oseen reminiscent vortex structures
embedded in wall-bounded flow has been reported in various
experiments and numerical simulations (see \eg Liu \etal\, 1996).
For the Lamb-Oseen vortex, the vorticity is non-zero only for the
axial component as (see figure \ref{fig:helix}($a$))
\begin{subeqnarray} \label{eqn:helisymvortLO}
\gdef\thesubequation{\theequation \textit{a--c}} \omega_r=0; \quad
\omega_{\theta} = 0; \quad \omega_z = \frac{\Gamma}{\pi
\varepsilon^2} \exp \Big (-\frac{r^2}{\varepsilon^2} \Big ).
\end{subeqnarray}

A more general model is the Batchelor vortex \cite[]{Batchelor1964},
which includes the non-uniform axisymmetrical axial velocity
distribution $u_z$ which approaches the Lamb-Oseen vortex in the
extreme. This vortex model is commonly used in instability studies
of swirling flows \cite[see][and references
therein]{HeatonPeake2007}. To describe experimental swirl flows
\cite[][Alekseenko \etal\, 1999]{Leibovich1978,Escudier1988}, the
Batchelor vortex model is usually referred to in the form
\begin{subeqnarray}\label{eqn:empirical}
\gdef\thesubequation{\theequation \textit{a--b}} u_{\theta} =
\frac{K}{r} \Big ( 1 - \exp(-\alpha r^2) \Big ); \quad u_z = W_1 +
W_2 \exp(-\alpha r^2)
\end{subeqnarray}

\noindent where $K$, $W_1$, $W_2$ and $\alpha$ are empirical
constants with simple physical interpretations as identified by
\cite{Okulov1996}
\begin{subeqnarray}
\gdef\thesubequation{\theequation \textit{a--d}} \Gamma = 2 \pi K;
\quad l=K/W_2; \quad u_0=W_1+W_2 \quad \mathrm{and} \quad
\varepsilon=1/\sqrt{\alpha}
\end{subeqnarray}

\noindent where $\Gamma$ is the vortex strength (circulation), $l$
is the pitch of the helical vortex lines, $u_0$ is the advection
velocity of the vortex and $\varepsilon$ is the effective size of
the vortex core with Gaussian axial vorticity distribution, see
figure \ref{fig:helix}($b$). The profiles given in
(\ref{eqn:empirical}) can reproduce experimentally determined swirl
flow with high accuracy. One possible approach is to test if the
empirical model (\ref{eqn:empirical}) can describe the longitudinal
vortex in the present case. However, in accordance with
\cite{Pierrehumbert1980} one needs to account for the possible
disturbance of the mirror vortex, resulting from the presence of the
wall. Another more suitable approach is therefore to extend the
Batchelor vortex model to model the flow by helical symmetry of the
vorticity, leaving no restrictions on the shape of the vortex core.
Flows with helical vorticity can be described by correlation between
the axial and circumferential vorticity vector components
\begin{subeqnarray} \label{eqn:helisymvort}
\gdef\thesubequation{\theequation \textit{a--c}} \omega_r=0; \quad
\omega_{\theta} = r \omega_z / l; \quad \omega_z = \frac{\Gamma}{\pi
\varepsilon^2} \exp \Big (-\frac{r^2}{\varepsilon^2} \Big )
\end{subeqnarray}

\noindent with the vorticity vector always directed along the
tangent of the helical lines, $x=r \cos \theta;$ $y=r \sin \theta;$
$z=l \theta$. Flows with helical vorticity can in addition be
characterized by the following condition for the velocity field
$\vec{u} = \{u_r ,u_{\theta} , u_z \}$.
\begin{subeqnarray} \label{eqn:helisym}
\gdef\thesubequation{\theequation \textit{a--b}}
u_z+\frac{r}{l}u_{\theta}=u_0 \equiv \mathrm{const.} \quad
\mathrm{or} \quad u_z=u_0-\frac{r}{l}u_{\theta}
\end{subeqnarray}

It can be shown that conditions (\ref{eqn:helisymvort}\textit{a--b})
and (\ref{eqn:helisym}) are equivalent \cite[see
e.g.][]{Okulov2004}. For a flow fulfilling the requirement of
(\ref{eqn:helisym}), the main flow parameters are $u_0$ and $l$.
Sometimes $u_0$, $u_z$ and $u_{\theta}$ are found directly from
measurements. The pitch $l$ can then be deduced from
(\ref{eqn:helisym}), but this approach might lead to an estimate of
high relative error if $u_z-u_0$ is small. Multiplying
(\ref{eqn:helisym}) by $u_z$ and integrating over the cross-section
of the flow one can obtain the pitch through the swirl number $S$
(Alekseenko \etal\, 1999)
\begin{equation}\label{eqn:pitchswirl}
l = - F_{mm}/(F_m-u_0 G),
\end{equation}

\noindent where $F_{mm}=\int_{\Sigma} \rho u_{\theta} u_z r \,
\mathrm{d}\Sigma$ is the angular momentum flux in the axial
direction, $F_{m}=\int_{\Sigma} \rho u_z^2 \, \mathrm{d}\Sigma$ the
momentum flux in the axial direction, $G$ the flow rate, $\rho$ the
fluid density and $\Sigma$ the cross-section area. All parameters
can now be determined: $u_0$ is found directly from the
measurements, $l$ is found through (\ref{eqn:pitchswirl}) and the
circulation $\Gamma$ and the vortex size $\varepsilon$ can be
extracted from (\ref{eqn:helisymvort}\textit{c}). Based on the
experimental observation the simple Batchelor vortex model is chosen
as

\begin{subeqnarray} \label{eqn:Lambaxial}
\gdef\thesubequation{\theequation \textit{a--b}} u_{\theta}=
\frac{\Gamma}{2 \pi r} \Big [ 1-\exp \Big ( -
\frac{r^2}{\varepsilon^2} \Big ) \Big ]; \quad  u_{z} = u_0 -
\frac{\Gamma}{2 \pi l} \Big [ 1-\exp \Big ( -
\frac{r^2}{\varepsilon^2} \Big ) \Big ].
\end{subeqnarray}

The only requirements of this simple model are the size of the
vortex core, the circulation, the helical pitch and the vortex
advection velocity.


\section{Testing of helical symmetry and embedded columnar vortex flow}\label{sec:results}
\begin{figure}
\begin{minipage}{0.5\linewidth}\vspace{-0.20cm}
\includegraphics[width=5.5cm]{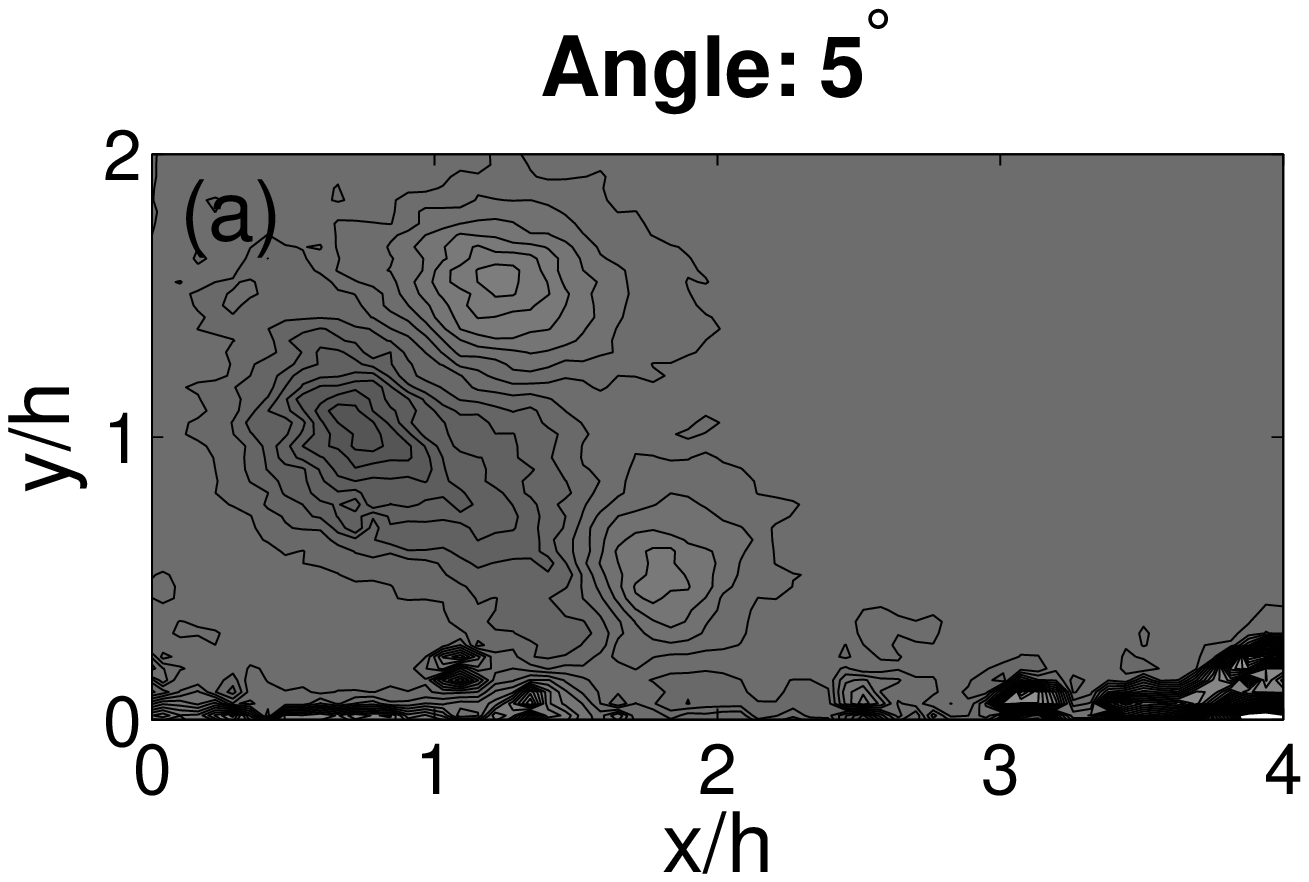}\vspace{-0.55cm}
\includegraphics[width=5.5cm]{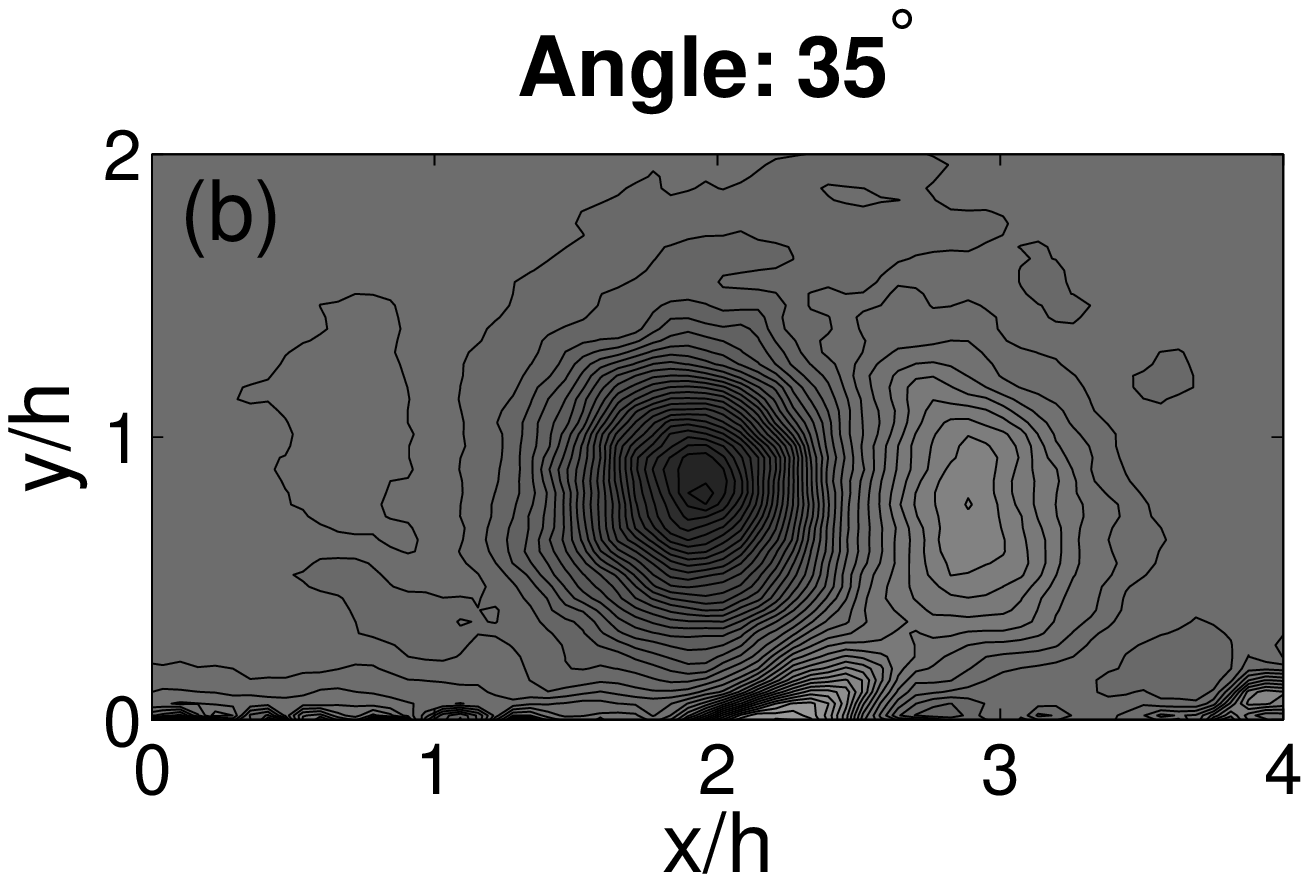}\vspace{-0.55cm}
\includegraphics[width=5.5cm]{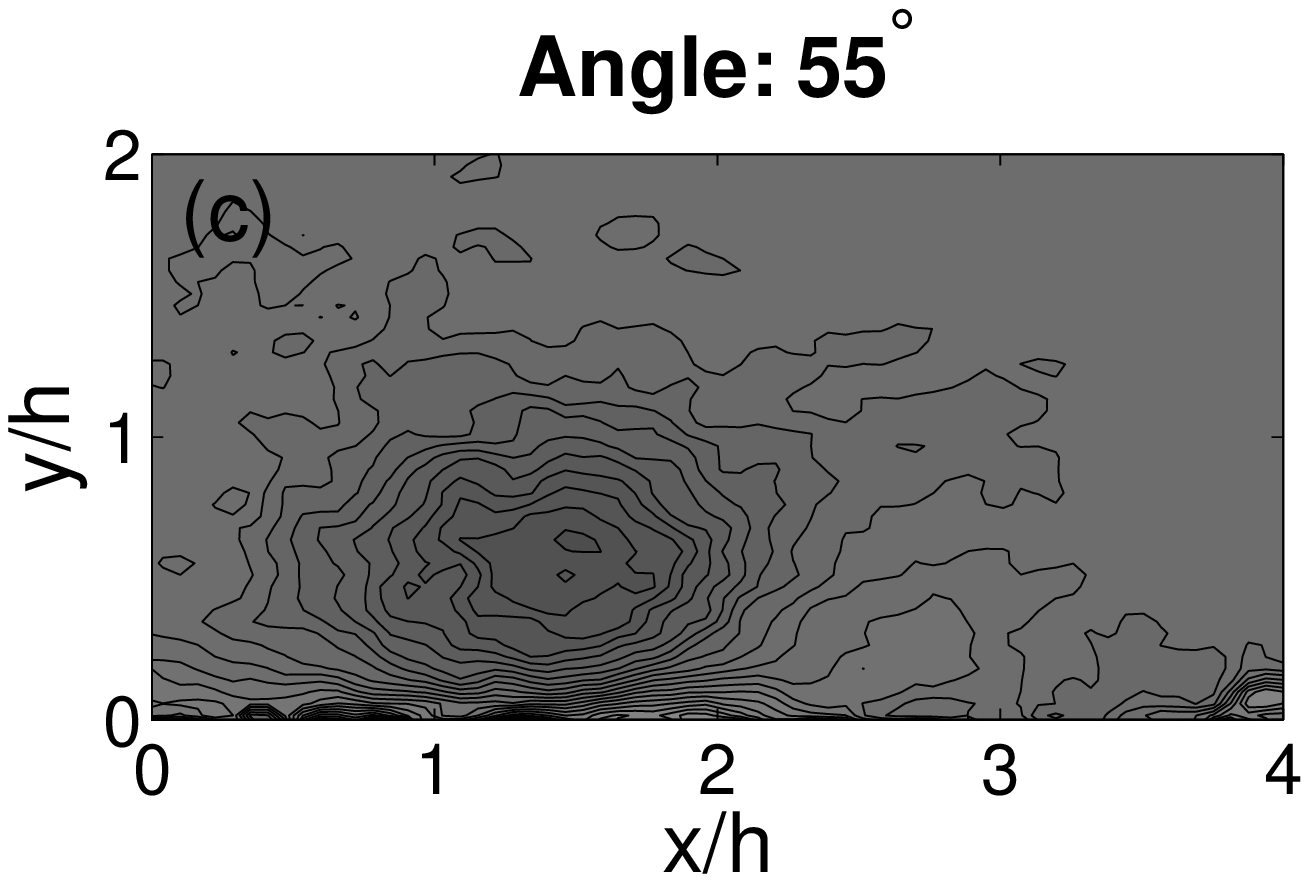}\vspace{-0.55cm}
\end{minipage}
\begin{minipage}{0.5\linewidth}\hspace{-1.7cm}
\includegraphics[width=10.5cm]{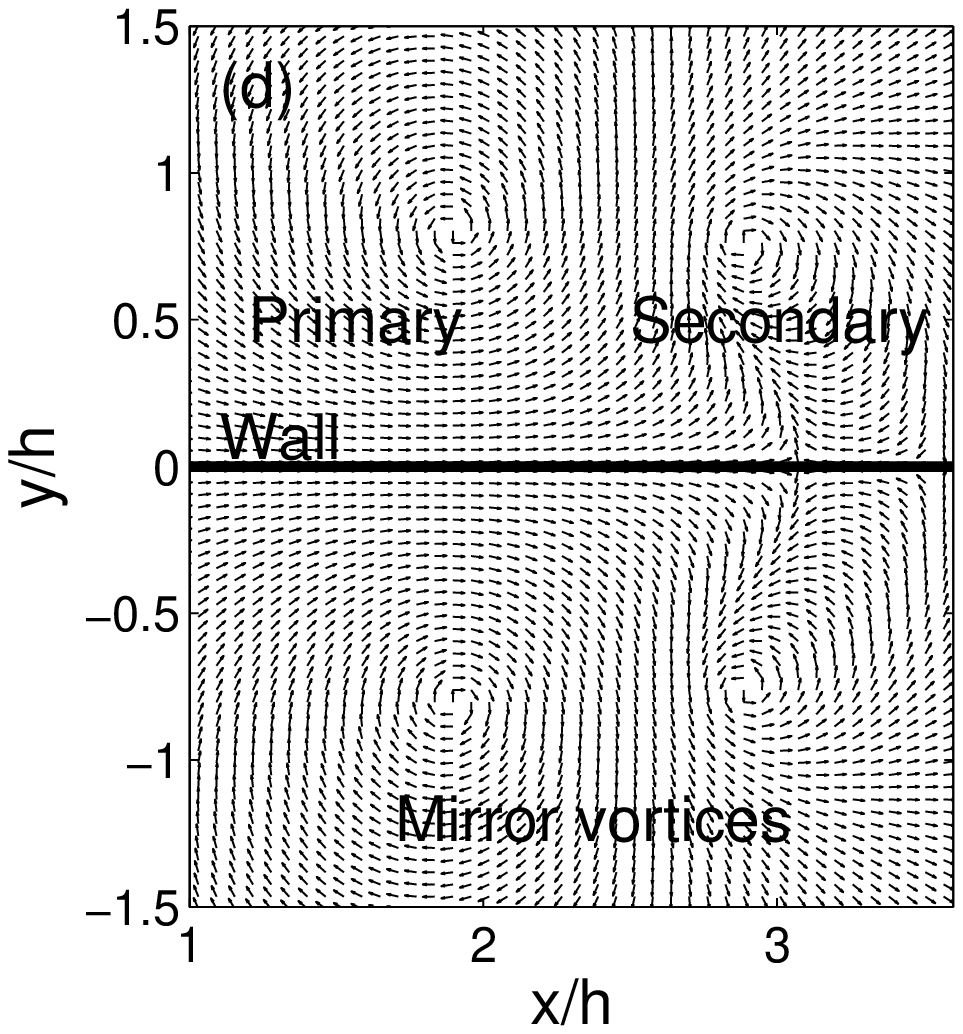}
\end{minipage}
\caption{Iso-contour maps of axial vorticity for device angles
(\textit{a}) $\beta =5^{\circ}$, (\textit{b}) $\beta =35^{\circ}$
and (\textit{c}) $\beta =55^{\circ}$. In (\textit{d}) a sketch
showing a sample velocity distribution of the primary and secondary
(upper half) and mirror vortices (lower half) is presented. The wall
is illustrated by a thick line at $y/h = 0$.} \label{fig:vortfields}
\end{figure}

\begin{figure}
\begin{minipage}{0.5\linewidth}
\includegraphics[width=7.0cm]{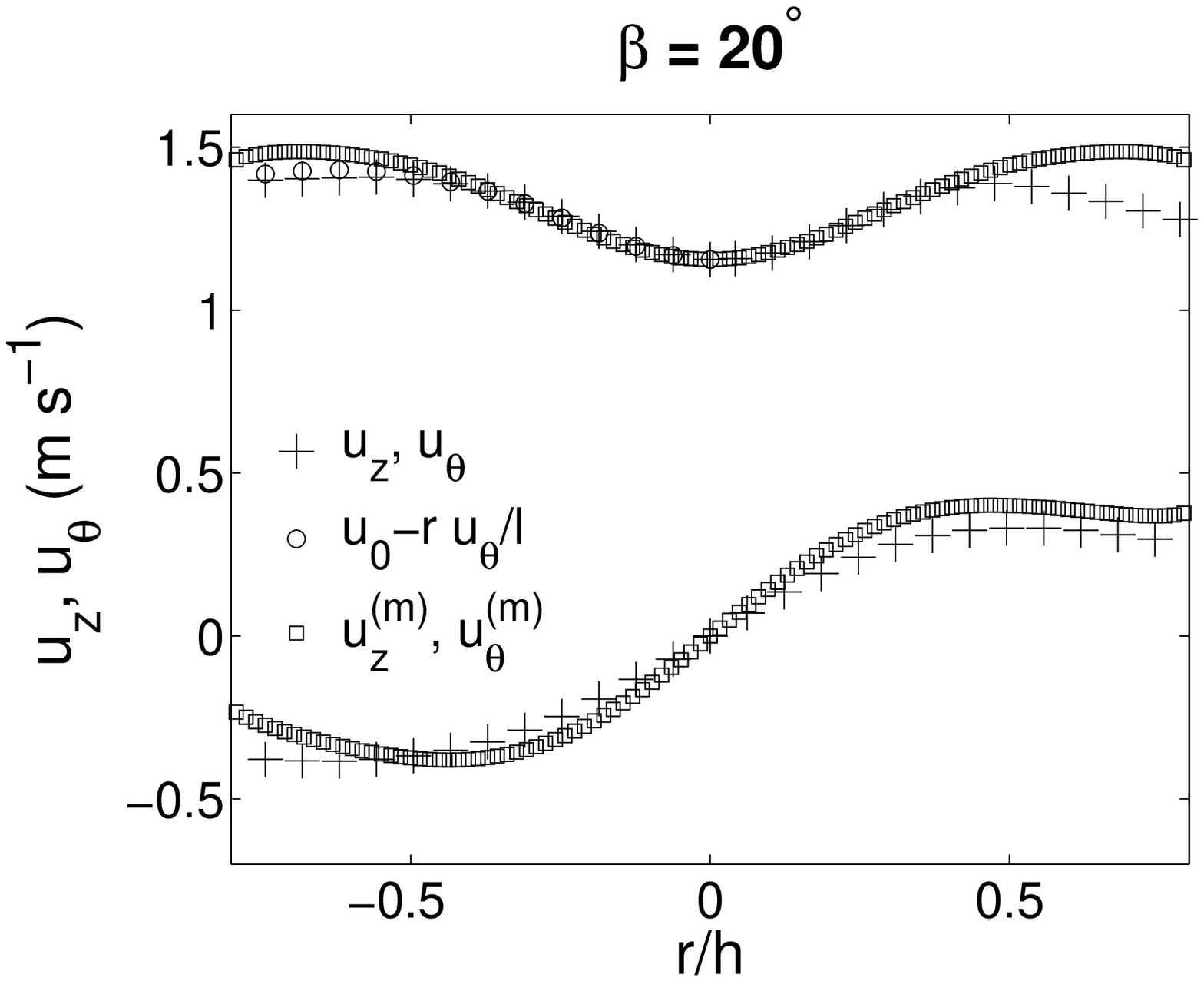}
\end{minipage}
\begin{minipage}{0.5\linewidth}
\includegraphics[width=7.0cm]{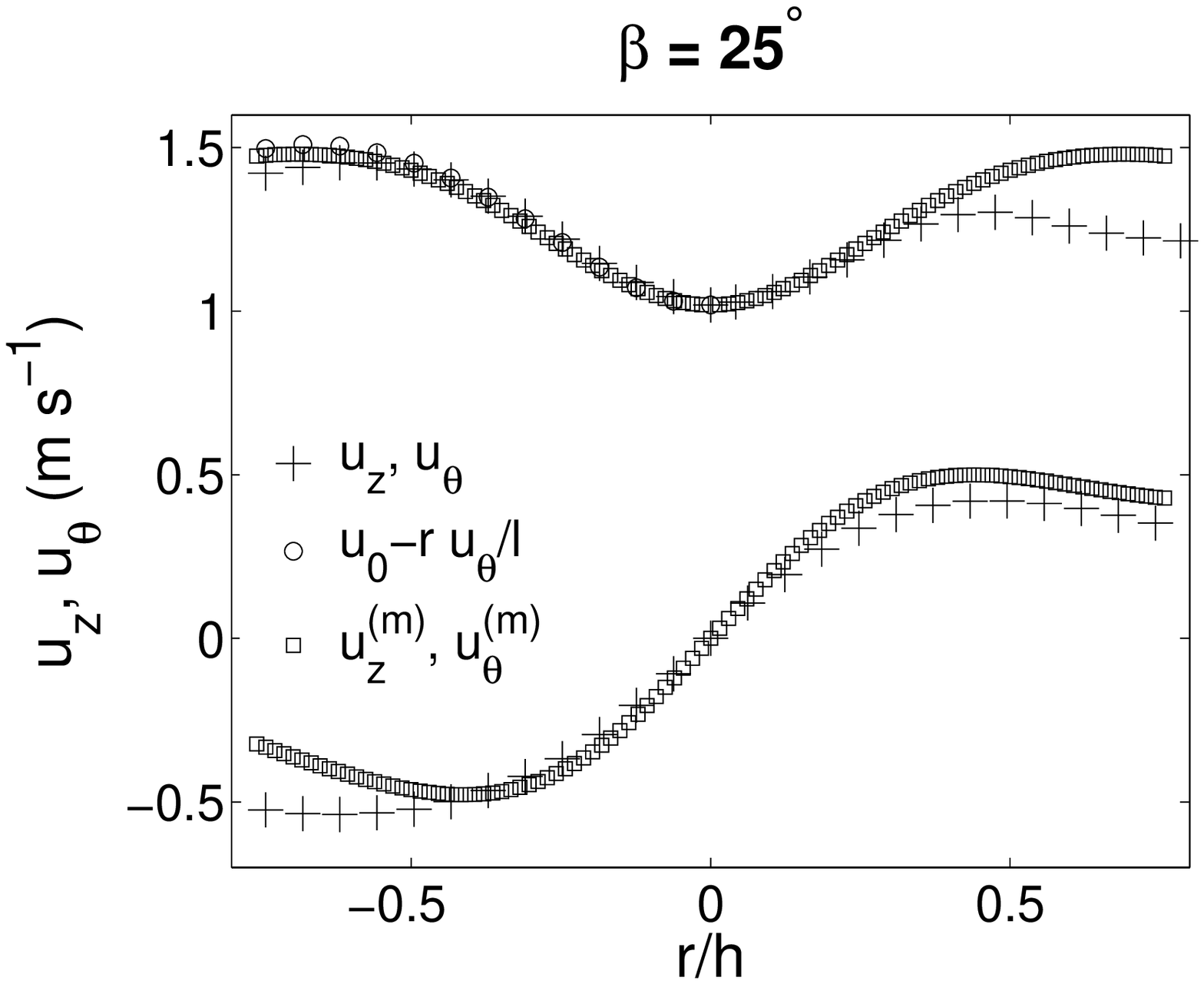}
\end{minipage}\\
\begin{minipage}{0.5\linewidth}
\includegraphics[width=7.0cm]{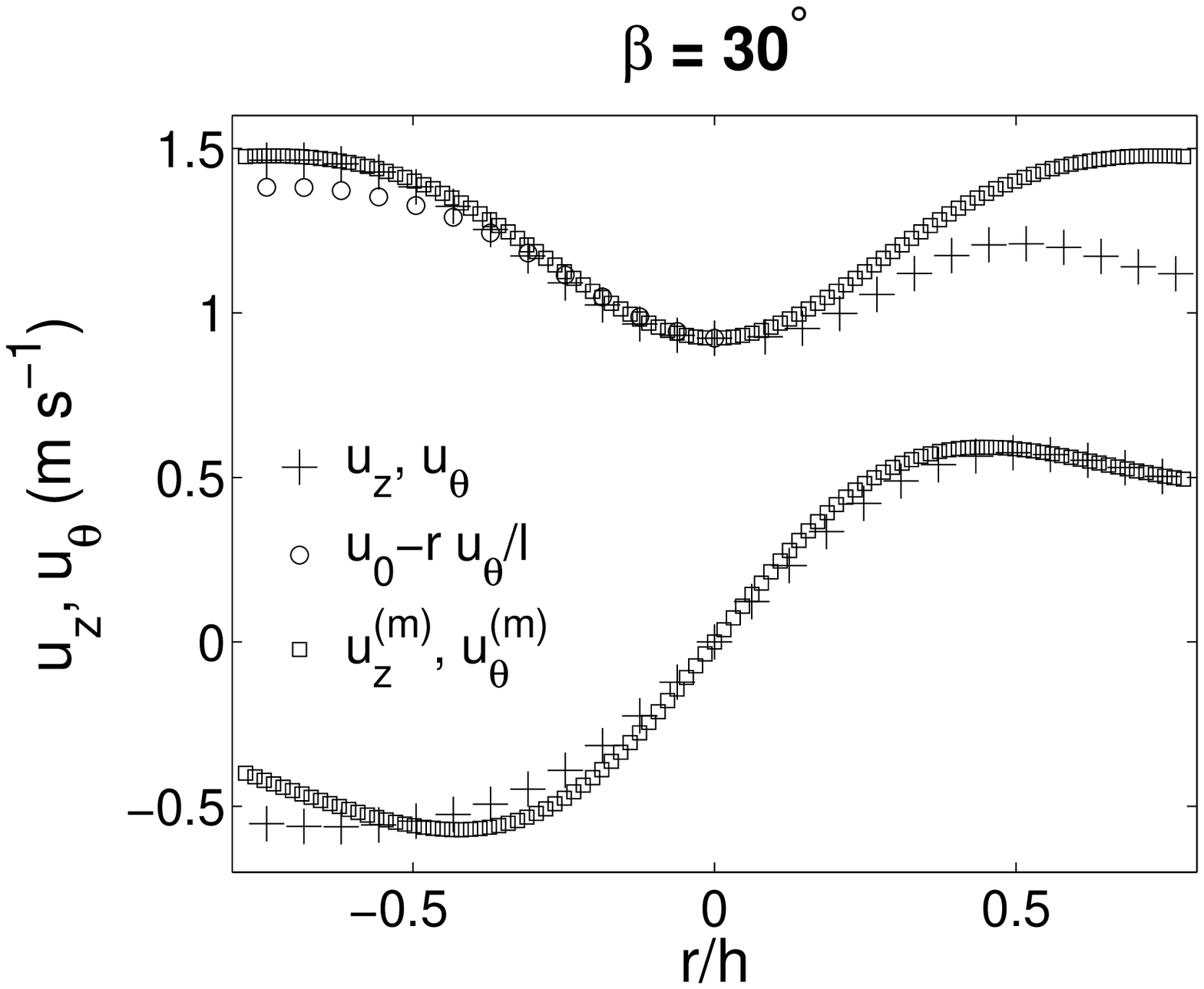}
\end{minipage}
\begin{minipage}{0.5\linewidth}
\includegraphics[width=7.0cm]{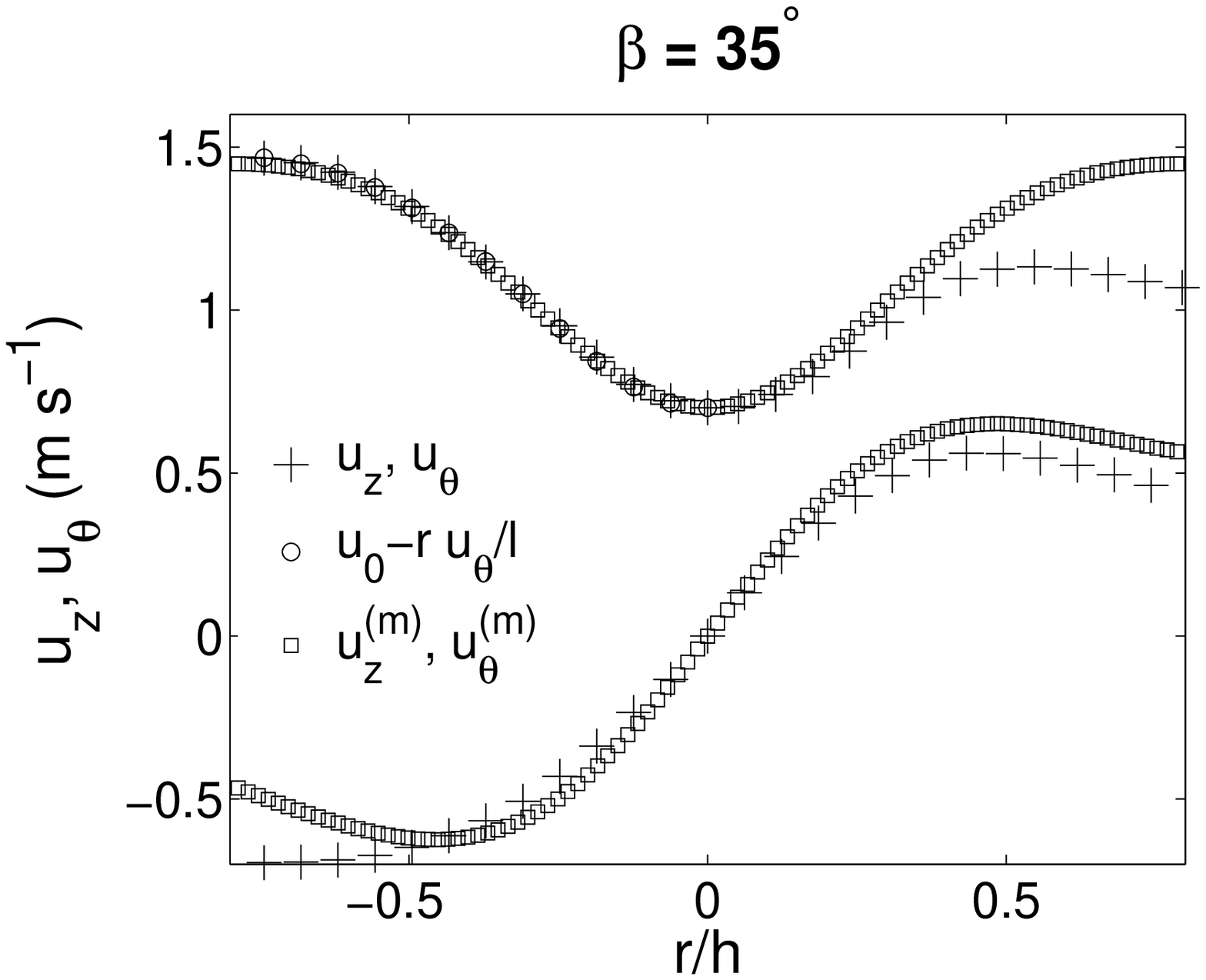}
\end{minipage}
\caption{Testing of helical symmetry of embedded vortices generated
by a vortex generator for various device angles $\beta$. The
measured axial ($u_z$, upper) and azimuthal ($u_{\theta}$, lower)
velocity profiles ($+$) are plotted. The measured values $u_z$ are
compared to the right hand side of (\ref{eqn:helisym}$b$) calculated
using the measured values $u_{\theta}$ ($\circ$). These computed
values are only displayed on the left side, since the flow on the
right side is perturbed by the secondary vortex. The two datasets
overlap quite well and the difference between the calculated and
measured values is hardly visible for some angles. Also displayed
are the azimuthal and axial velocity profiles of the utilized vortex
model (\ref{eqn:Lambaxial}\textit{a--b}) ($\square$).}
\label{fig:helcheck}
\end{figure}

The analysis of the embedded vortices was done based on the ensemble
averaged complete cross-plane velocity field from the SPIV
measurements and the therefrom derived axial vorticity component,
see figure \ref{fig:vortfields}($a$--$c$). The iso-contour maps of
axial vorticity reveal the presence of a secondary vortex, which can
be seen next to the main vortex at $x/h \approx 3$ in the
iso-contour map for $\beta = 35 ^{\circ}$ in figure
\ref{fig:vortfields}($b$). Figure \ref{fig:vortfields}($d$) displays
a sketch of the primary and the secondary vortices in the upper half
and the mirrored velocity field in the lower half.

Figure \ref{fig:helcheck} shows the measured axial $u_z$ (upper) and
azimuthal $u_{\theta}$ (lower) velocity profiles ($+$) for various
values of the device angle $\beta$ extracted along a line parallel
to the wall through the centre of the primary vortex. Verification
of the hypothesis of helical symmetry was done by comparing the left
($+$) and right ($\circ$) hand side of (\ref{eqn:helisym}$b$)
calculated from the measured values of $u_z$ and $u_{\theta}$. The
helical pitch $l$ was found by minimizing the sum of the residuals
of the right and left hand side of (\ref{eqn:helisym}$b$) in a least
squares sense for a limited set of points in the radial direction.
The values computed from the right hand side ($\circ$) are only
displayed on the left side of the primary vortex centre, since the
flow on the right side is perturbed by the secondary vortex. The two
datasets overlap quite well, which is why the difference between the
calculated and measured values is hardly visible for some angles.

The axial vorticity fields of the vortices derived from the
measurement data have Gaussian distributions and one can therefore
use (\ref{eqn:helisymvort}\textit{c}) to find the circulation
$\Gamma$ and vortex size $\varepsilon$ of both the main and the
secondary vortices. The local flow characteristic $u_0$ was found
directly from the measurements and the helical pitch $l$ was
obtained from (\ref{eqn:pitchswirl}), yielding a result which agreed
well with the values obtained by minimizing the sum of residuals of
(\ref{eqn:helisym}$b$) in a least squares sense. The azimuthal
($u_{\theta}^{(m)}$) and axial ($u_z^{(m)}$) velocities induced by
the main vortex were modelled using
(\ref{eqn:Lambaxial}\textit{a--b}) ($\square$) and should be
compared to the measurements ($+$), see figure \ref{fig:helcheck}.
This simple model is decoupled from all additional flow effects such
as the secondary and mirror vortices and the non-uniform flow due to
the presence of the wall. In spite of this, the model describes the
primary vortex flow well in the regime under consideration.

The secondary vortex is present with varying strength at all
considered device angles, introducing a disturbance in the flow
field of the main vortex and thereby causing asymmetry. The mirror
vortices will have the same effect on the symmetry of the main
vortex. For angles smaller than $15^{\circ}$, an additional vortex
was observed, increasing the complexity of the flow by yielding a
three vortex system perturbing the vorticity distribution and the
velocity field considerably, see figure \ref{fig:vortfields}($a$).
For small values of $\beta$, the vortex system becomes more
complicated and equation (\ref{eqn:helisymvort}\textit{c}) is not
representative for the actual flow. For angles larger than
$40^{\circ}$, the fit again becomes worse due to the instabilities
for high values of circulation at large device angles, see figure
\ref{fig:vortfields}($c$). For increasing values of $\beta$, the
vorticity component will surpass from streamwise to more and more
spanwise, eventually resulting in pure shedding in the extreme
$\beta=90^{\circ}$. Due to the decreasing longitudinal vorticity
component for large values of $\beta$, the helical symmetry is
destroyed. The deviations arise because we have a simple model with
linear interactions, which is being compared to measured values
originating from a more complex representation of the flow.
Nonlinearities are not captured by the linear model and become
increasingly dominant outside of the range $20^{\circ} \leq \beta
\leq 40^{\circ}$.

\begin{figure}
\begin{minipage}{16pc}
\includegraphics[width=7.35cm]{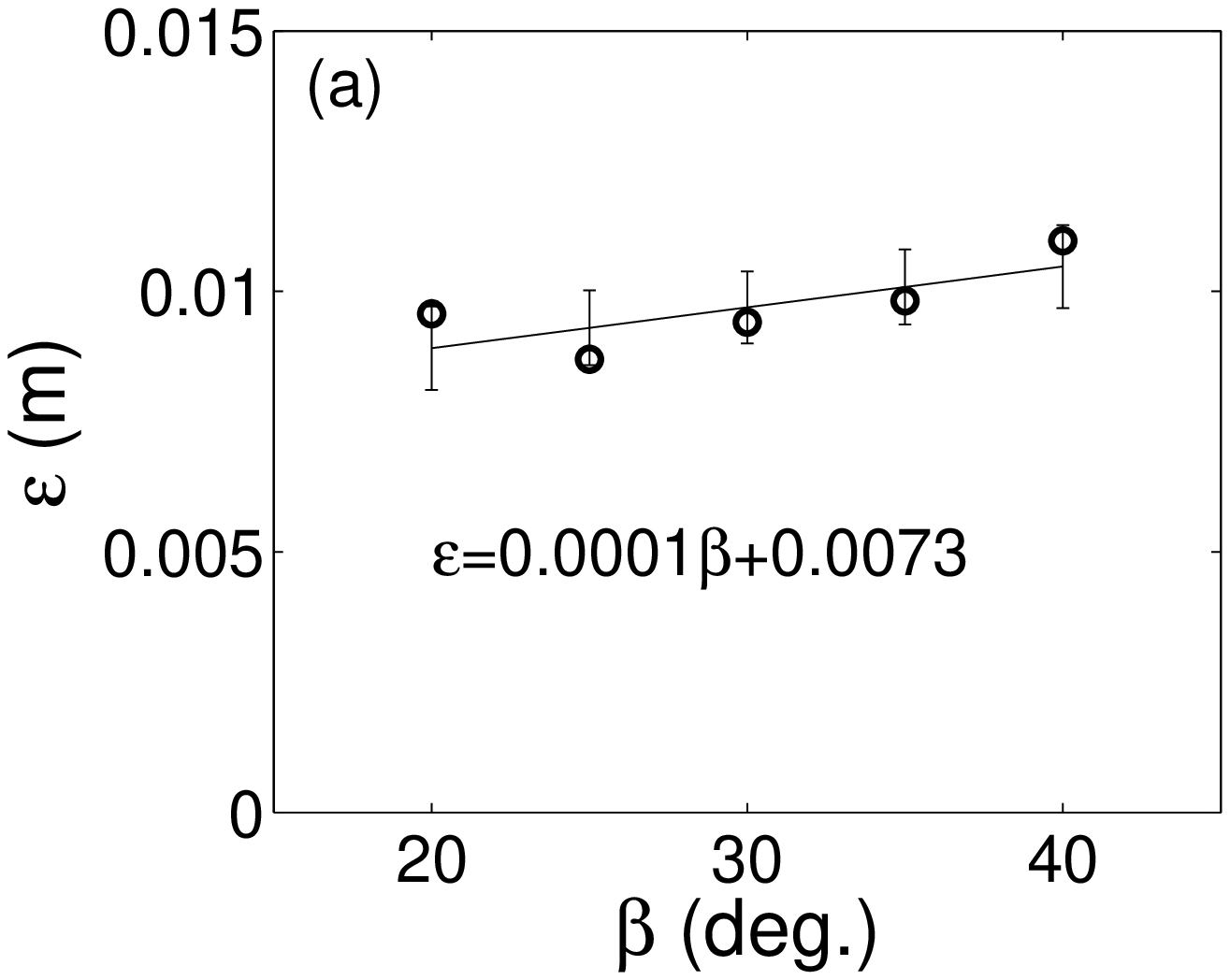}
\end{minipage}
\begin{minipage}{16pc}
\includegraphics[width=7.35cm]{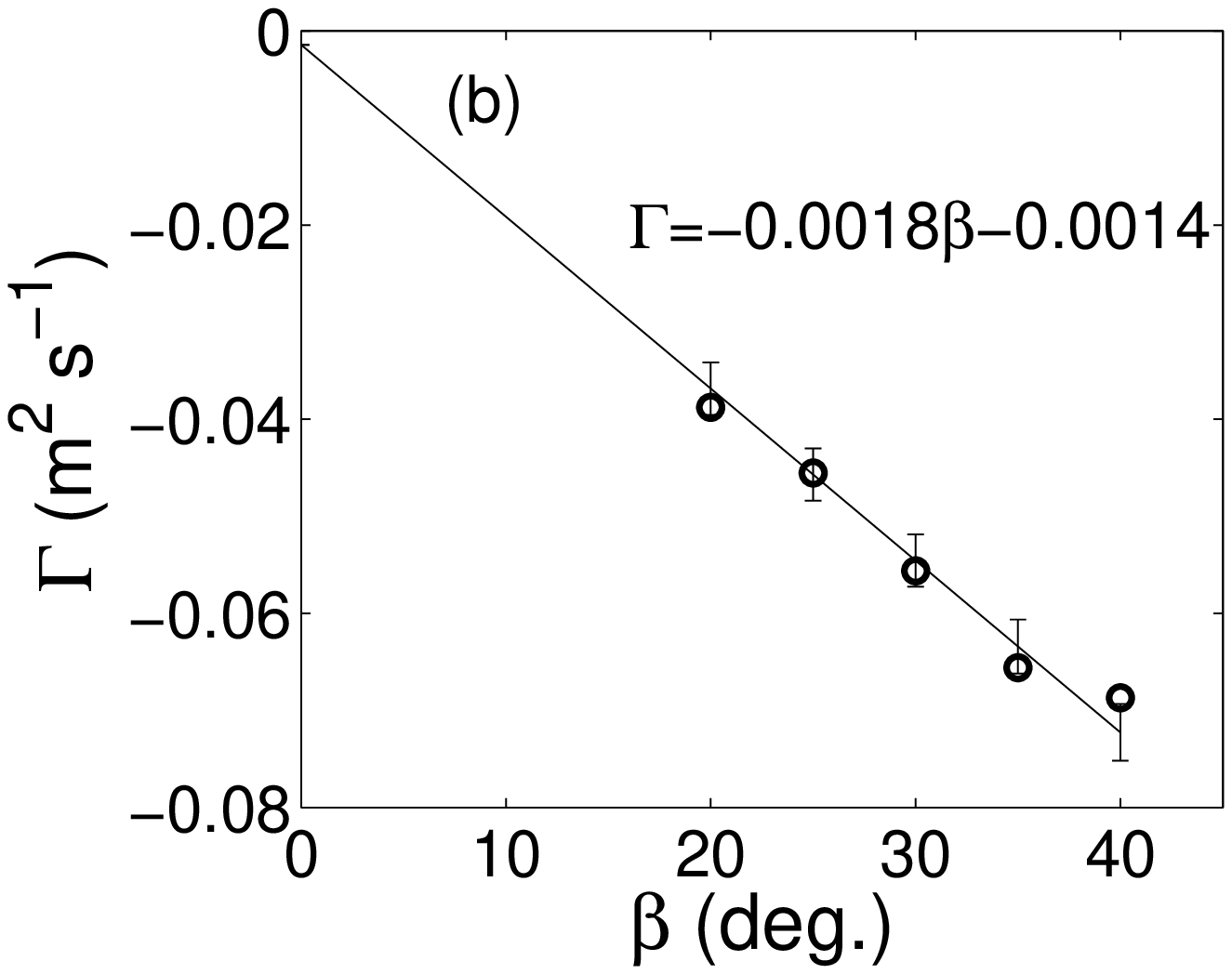}
\end{minipage}
\begin{minipage}{16pc}
\includegraphics[width=7.35cm]{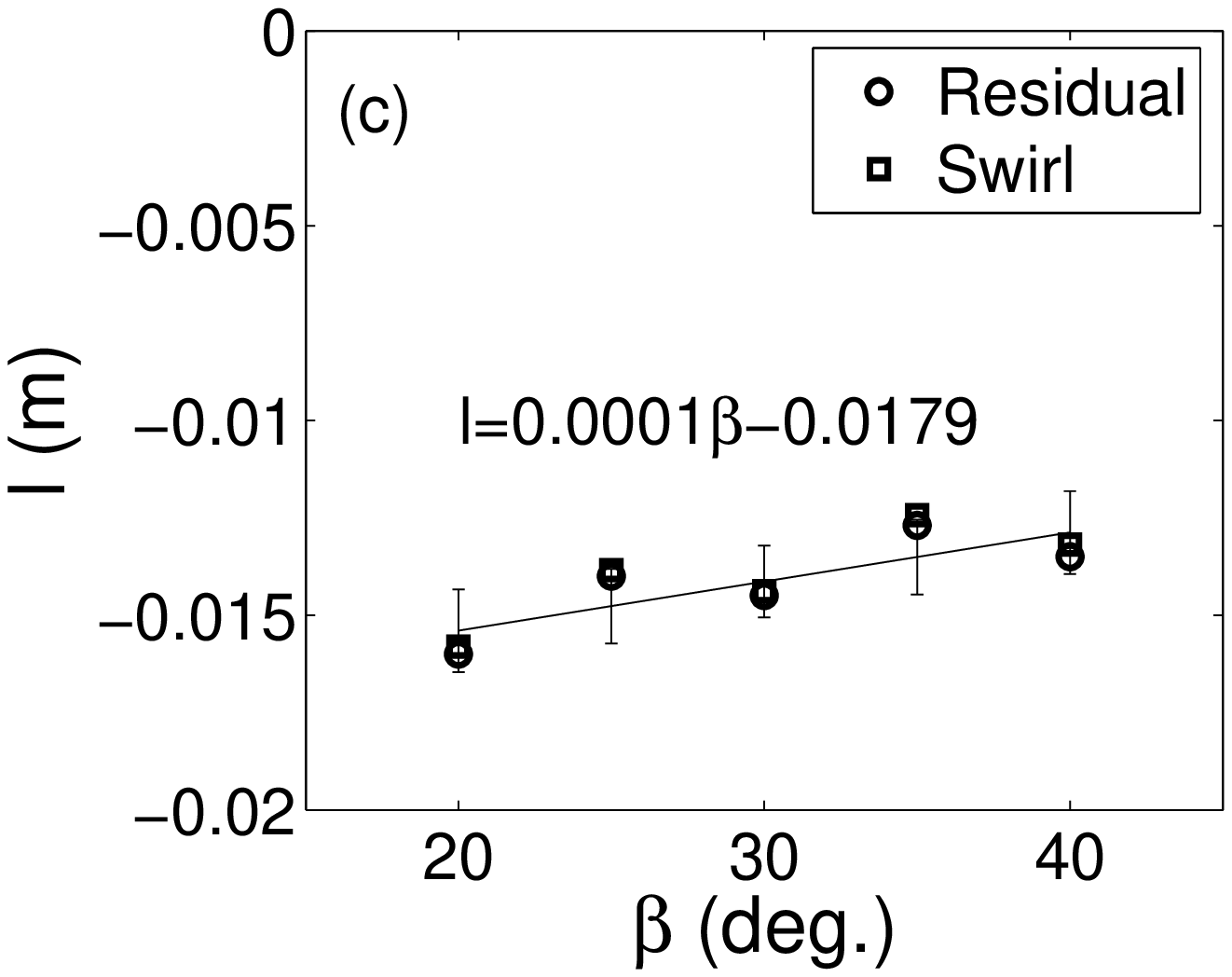}
\end{minipage}
\begin{minipage}{16pc}
\includegraphics[width=7.35cm]{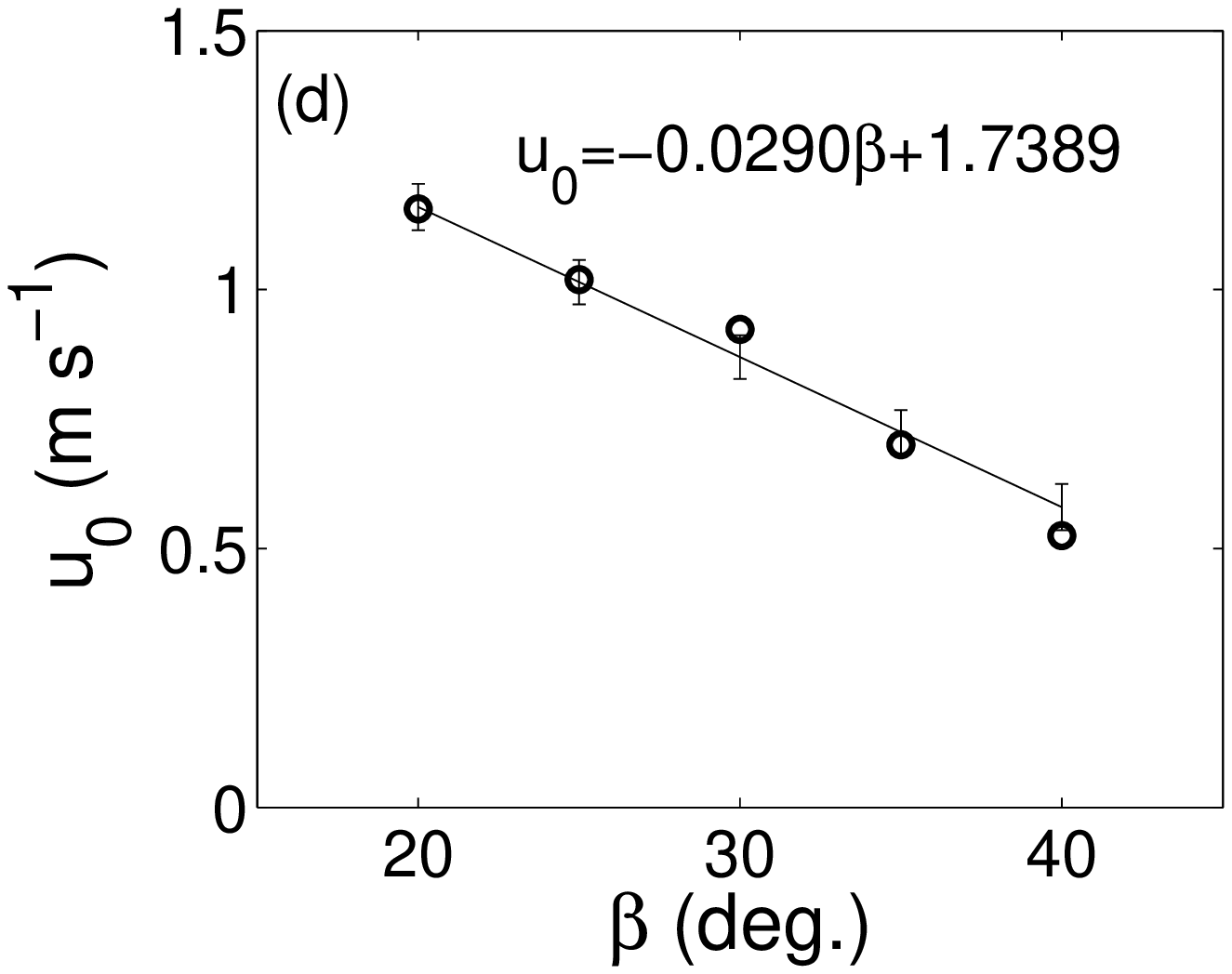}
\end{minipage}
\caption{Device angle $\beta$ dependency of (a) the vortex radius
$\varepsilon$, (b) circulation $\Gamma$, (c) helical pitch $l$ and
(d) advection velocity of the vortex $u_0$. The datasets are
provided with a linear fit in a least squares sense with
corresponding error bars. In (c), the helical pitch $l$ has been
obtained from minimizing the residual of
(\ref{eqn:helisym}\textit{b}) in a least squares sense ($\circ$) and
is compared to the helical pitch of the swirling flow
(\ref{eqn:pitchswirl}) ($\square$) obtained using
(\ref{eqn:Lambaxial}\textit{a--b}).} \label{fig:params}
\end{figure}

Figure \ref{fig:params} shows the device angle dependency of the
parameters of the problem in the range $20^{\circ} \leq \beta \leq
40^{\circ}$. The device angle dependency of the vortex radius
$\varepsilon$ and circulation $\Gamma$, obtained from the Gaussian
fit (\ref{eqn:helisymvort}\textit{c}) of the vorticity, are shown in
figures \ref{fig:params}($a$) and \ref{fig:params}($b$)
respectively. The datasets have been fitted with a linear
approximation in a least squares sense, with corresponding error
estimates. For the circulation, the fitting has been extrapolated to
zero device angle. As expected, the extrapolated circulation is
approximately zero at $\beta = 0^{\circ}$, since a device with no
angle to the flow ideally will not give rise to any circulation. One
can see that the vortex size and the magnitude of the circulation
increase linearly with the device angle. The device angle dependency
of the helical flow characteristics $l$ and $u_0$ with linear
fitting and error bars are shown in figures
\ref{fig:params}(\textit{c}) and \ref{fig:params}(\textit{d})
respectively. In figure \ref{fig:params}(\textit{c}), the values of
the helical pitch $l$ obtained from minimizing the residual of
(\ref{eqn:helisym}\textit{b}) in a least squares sense ($\circ$) are
compared to the helical pitch of the swirling flow
(\ref{eqn:pitchswirl}) ($\square$) obtained using the axial and
azimuthal velocities of the model
(\ref{eqn:Lambaxial}\textit{a--b}). These two datasets are strongly
correlated and it is also seen that the pitch only varies marginally
with device angle $\beta$. The advection velocity of the vortex
$u_0$ decreases linearly with device angle.


\section{Conclusions}\label{sec:concl}

Vortices generated by a passive rectangular vane-type vortex
generator of the same height as the boundary layer thickness in a
flat plate wall-bounded flow have been studied experimentally. It
has been shown that the embedded vortices possess helical symmetry
in the device angle range $20^{\circ} \leq \beta \leq 40^{\circ}$.
The flow field in the considered regime consists of two vortices,
the primary one and a secondary one. Outside of this range
additional flow effects influence the helical vortex in a
destructive way, deterring the helical symmetry to persist. The
vorticity distribution across the vortices is Gaussian, yielding
estimates of the vortex radius $\varepsilon$ and circulation
$\Gamma$ through (\ref{eqn:helisymvort}\textit{c}). This rendered
the possibility to describe the flow in a realistic and simple
fashion, utilizing a model for the azimuthal and axial velocity
components, (\ref{eqn:Lambaxial}\textit{a--b}). Comparison of these
modelled velocities to the measured data showed to concur well in
the device angle regime under consideration. Being the main flow
characteristics of a vortex with helical symmetry, the determination
of the helical pitch $l$ and the axial velocity at the vortex centre
$u_0$ is of great importance to characterize the vortex
(\ref{eqn:helisym}\textit{a--b}). $u_0$ was obtained directly from
the measurements, whereas the pitch was determined by minimization
of the sum of the residuals of (\ref{eqn:helisym}\textit{b}) in a
least squares sense or alternatively from (\ref{eqn:pitchswirl}).
The results of these two methods for pitch evaluation showed a high
degree of concurrence.

The vortex radius $\varepsilon$, the circulation $\Gamma$, the
helical pitch $l$ and the advection motion of the vortex (or axial
velocity at the vortex centre) $u_0$ all showed linear dependency
with the device angle $\beta$. These simple relations render it
possible to predict these parameter values for device angles in the
range $20^{\circ} \leq \beta \leq 40^{\circ}$ well and thereby
determine vortex strength, size and axial flow distribution. They
also facilitate theoretical studies analysing \eg stability and aid
in modelling the flow within this range. The vortex radius showed a
weak increase with increased device angle $\beta$, while the
circulation $\Gamma$ showed a large increase in magnitude. The
vortex advection velocity $u_0$ decreased with increased device
angle while the helical pitch did not change notably and can, for
the purpose of the model, be considered close to constant.

\begin{acknowledgments}
The Danish Research Council, DSF, is acknowledged for their
financial support of the project under grant number 2104-04-0020.
\end{acknowledgments}

\bibliographystyle{jfm}

\end{document}